\documentclass[journal]{IEEEtran}
\pdfoutput=1
\usepackage{cite}
\usepackage{amsfonts}
\usepackage{algorithm}
\usepackage{booktabs}
\usepackage{array}
\usepackage{multirow}
\usepackage{graphicx}
\usepackage{subfig}
\usepackage{listings}
\usepackage[utf8]{inputenc}
\usepackage[T1]{fontenc}
\usepackage{url}
\usepackage{hyperref}

\begin{document}
%\bstctlcite{IEEEexample:BSTcontrol}

\title{Using Multi-Agent Reinforcement Learning in Auction Simulations}

%\author{\IEEEauthorblockN{Medet Kanmaz}
%\IEEEauthorblockA{\textit{Graduate School of Informatics} \\
%\textit{Middle East Technical University} \\
%Ankara, Turkey \\
%medet.kanmaz@metu.edu.tr}
%\and 
%\newline
%\IEEEauthorblockN{Elif Surer}
%\IEEEauthorblockA{\textit{Graduate School of Informatics} \\
%\textit{Middle East Technical University} \\
%Ankara, Turkey \\
%elifs@metu.edu.tr}
%}

\author{Medet~Kanmaz and Elif~Surer\\
Department of Modeling and Simulation, Graduate School of Informatics, Middle East Technical University, Ankara, Turkey\\
{\tt\small medet.kanmaz@metu.edu.tr, elifs@metu.edu.tr}

}

\maketitle

\begin{abstract}
Game theory has been developed by scientists as a theory of strategic interaction among players who are supposed to be perfectly rational. These strategic interactions might have been presented in an auction, a business negotiation, a chess game, or even in a political conflict aroused between different agents. In this study, the strategic (rational) agents created by reinforcement learning algorithms are supposed to be bidder agents in various types of auction mechanisms such as British Auction, Sealed Bid Auction, and Vickrey Auction designs. Next, the equilibrium points determined by the agents are compared with the outcomes of the Nash equilibrium points for these environments. The bidding strategy of the agents is analyzed in terms of individual rationality, truthfulness (strategy-proof), and computational efficiency. The results show that using a multi-agent reinforcement learning strategy improves the outcomes of the auction simulations.
\end{abstract}

\section{Introduction}
A multi-user game is a setting in which one player’s payoff depends not only on that player’s action but also on what the other player is doing. In many games, players do not have a dominant strategy among a multitude of options. A well-balanced game provides meaningful options to its players while avoiding dominant strategies. On the other hand, in real life, players might have dominant strategies —strategies that represent the best response to all possible strategies. The best response, on the other hand, is the best strategy given based on what you think or believe the other competitor might do. Nash equilibrium is a kind of stability, described as a combination of strategies in a game so that rational players do not have an incentive to deviate from their choice \cite{IEEEhowto:zamir}. A rational player in this context is more related to the selection of actions from a set of alternatives. If an agent is capable of ranking the outcomes of each possible action and able to calculate the possible outcomes for a given sequential set of actions, or able to choose the most preferred outcome with a given action of the other player, he or she is called a “rational” player. In addition to rationality, in the auction design, we assume that the actors are strategy-proof, which means that they play the game truthfully. Therefore, if there is a dominant strategy among a set of alternative actions, the player chooses that action or the best strategy truthfully.

In this paper, we create a Reinforcement Learning (RL) environment with a game design where the agents are competing to get a product in an auction and get positive or negative rewards based on the outcome of their actions. Each agent has a “private” value for the product. It is “private” since the counterpart has no idea about the agent’s appraisal for the product. The action set for the agents consists of “bid” and “stop.” When an auction ends, the reward for the winner is the difference between the winning price of the auction and the private value of the agent while the loser gets a negative reward, which represents the auction fee.

\section{Related Research} \label{sec:relatedresearch}

Multi-agent reinforcement learning (MARL) is an important and interdisciplinary research area that has been applied to several domains such as computer science, bioinformatics, and operations research \cite{IEEEhowto:sutton} \cite{IEEEhowto:stone}. Nanduri and Das \cite{IEEEhowto:nanduri} propose a stochastic approximation-based reinforcement learning algorithm to converge the Nash Equilibrium of n-player matrix games. The results of their algorithm on 16 different matrix games are compared to a software program, GAMBIT \cite{IEEEhowto:mckelvey}. Also, the authors propose a practical application on the electric power market to get strategic bidding. It is important to offer an actual use case for the RL agents. Similarly, Oroojlooyjadid et al. \cite{IEEEhowto:oroojlooyjadid} propose multi-agent learning solutions to real-life examples. The authors propose a machine learning algorithm based on deep Q-networks to optimize the replenishment decisions at a given stage for inventory optimization of a beer game.

The proposed examples are not restricted to games \cite{IEEEhowto:littman}\cite{IEEEhowto:bakker}\cite{IEEEhowto:crites}\cite{IEEEhowto:hsu}\cite{IEEEhowto:riedmiller}\cite{IEEEhowto:tesauro}. Nowe et al. \cite{IEEEhowto:nowe}, for example, focus on the application side of the reinforcement learning techniques in multi-agent reinforcement learning with the game theory perspective based on economic research. The authors point out that in the multi-agent setting, the agents should not have opposing goals to get a clear and optimal solution. Also, it is important to address all the issues regarding the dynamic environment. Since there is incomplete information about the environment, each agent observes the changing state rather than the other agent’s action, which is the reason for the reward of the agent. In that research, similar to those proposed beforehand, the RL agents have similar goals. In addition to the goals of the RL agents, efficiency in computation is an important detail. Focusing on practical learning of human behavior in zero-sum settings, Ling et al. \cite{IEEEhowto:ling} aim to prevent the limitations in the space of player strategies and create efficient and measurable computations. Therefore, they generate a more general model called “Nested Logit Quantal Response Equilibrium” in which the ideas in behavioral science are used, and the changes in the degree of player rationality in different stages of the game are allowed.

Searching for a stationary Nash equilibrium in a finite discounted general-sum stochastic game, Prasad et al. \cite{IEEEhowto:prasad} consider a Markov game, a finite stochastic game evolving over discrete time instants. They construct their model with the target of maximization of the expected discounted value of the agents’ rewards in a dependent transition dynamics and agent actions. The use of necessary and sufficient Stochastic Game-Sub-Problem (SG-SP) conditions lead them to develop converging two actor-critic algorithms for model-based and model-free, named as OFF-SGSP and ON-SGSP, respectively. ON-SGSP algorithm is evaluated to perform better than MARL algorithms, and its convergence is observed to be quicker. Lin et al. \cite{IEEEhowto:lin} focus on stochastic games, which are generalized forms of Markov decision processes to a game-theoretic scenario, and they use multi-agent inverse reinforcement learning (MIRL) constructed in this framework. The results of a competitive two agent zero-sum MIRL problem and its comparison with Bayesian MIRL in the setting of an abstract soccer game show that the covariance structure is relatively more important when the mean value in reward priors is considered.

Cai et al. \cite{IEEEhowto:cai} search an optimal model for the allocation of buyer impression to the potential sellers taking place in e-commerce websites. They use reinforcement mechanism design, and this general framework focuses on the strategic behaviors of the sellers and the history of impressions, prices, transactions, collected revenue, and the actions are evaluated under a Markov decision process. The design of the actor-critic policy gradient algorithm is generated under the setting of a Deep Deterministic Policy Gradient (DDPG) algorithm and an Impression Allocation (IA) algorithm performing the best outcome, is created.

\section{Methodology} \label{sec:methodology}

In this game design, the agents are competing to get a product P in an auction. Each agent has a private appraisal for the selling product. The action set for the agents consists of “bid a new price” (the price is an available option in the auction which may be increasing or decreasing for the different auction types) and “stop bidding.” When an auction ends, the reward for the winner is the difference between the winner price and the appraisal of the agent; the loser gets zero outcomes (or negative reward represents the auction fee). The figure below illustrates a simplistic case for an auction in reinforcement learning. As can be seen in Figure 1, the agents have the action space of “Bid” and “Stop.” Bidding means the initial action to step in the auction, as well as increasing the current price of the product. “Stop” action, on the other hand, has the meaning of resting the auction. Each agent has a private appraisal for the auctioned product. Depending on the auction type, the next state is determined by the bidding (i.e., actions were taken) behavior of the agents. The final reward is the difference between the private appraisal for the product and the close price of the auction. The reward will be positive when the price is set below the appraisal level; otherwise, it will be 0 or negative. In the case scenario that this study suggests, there is no entrance or bidding fee. Therefore, if an agent draws the auction in any stage, the reward will be zero.

\begin{figure}[!t]
\centering
\includegraphics[width=3.2in]{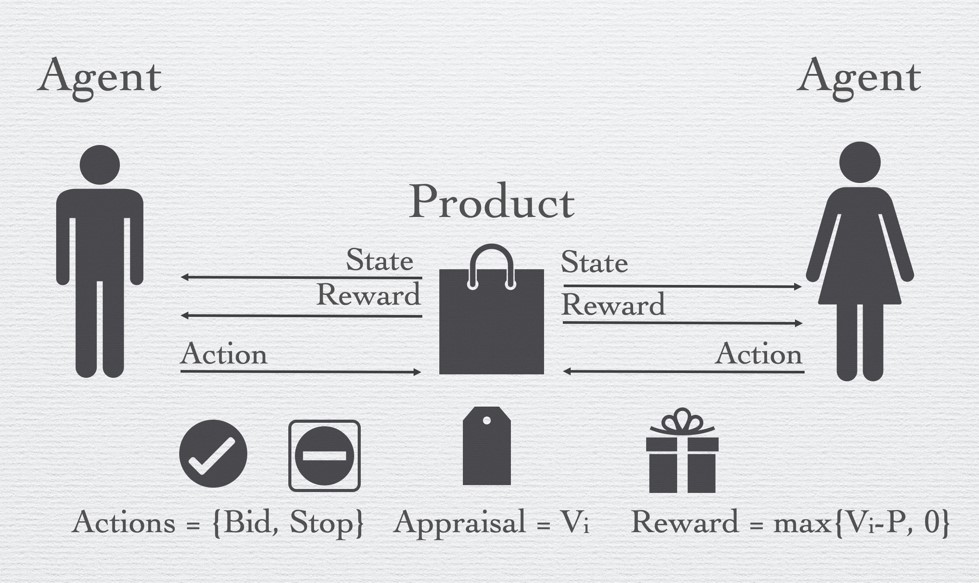}
% where an .eps filename suffix will be assumed under latex,
% and a .pdf suffix will be assumed for pdflatex; or what has been declared
\DeclareGraphicsExtensions.
\caption{A reinforcement learning mechanism for an auction design.}
\label{fig1}
\end{figure}

\section{Experiments}

The reinforcement learning agents are trained in an open bid with one period model initially. In this model, each agent has just one chance to bid, and the order of the bidding is sequential: the second agent sees the first agent’s price and decides either bidding or not and so on.

Figure 2 and Figure 3 show the hyperparameters, the heatmap, and the rewards for the agents in a bidding environment with a ceiling price of \$5, agent’s private valuations of \$3.5 with a non-divisible product and an entrance fee for \$1 (which means the losing party has a negative payoff). In this scenario, the bid increment is \$1, and the agents have to bid scaler bids. As can be seen in the figure, after 1000 epochs, the first agent’s action is saturated (a dominant strategy) to bid “\$4”, which results in a \$0.5 positive reward for the agent. On the other hand, the second agent has no dominant strategy but has a mixed strategy to bid either draw (0) or lower bid, which results in a “0” payoff for the second agent. As for the second agent, increasing the bid result in a higher current price (i.e., \$5), which results in a negative outcome (-\$1.5) for the second agent since the appraisal for the product is set as \$3.5. This is the equilibrium for the game, first agent bids “4” and gets “0.5” payoff, and the second agent draws the auction with a “-1” payoff.

\begin{figure}[htp]

\subfloat[Reward function for the first player.]{%
  \includegraphics[clip,width=\columnwidth]{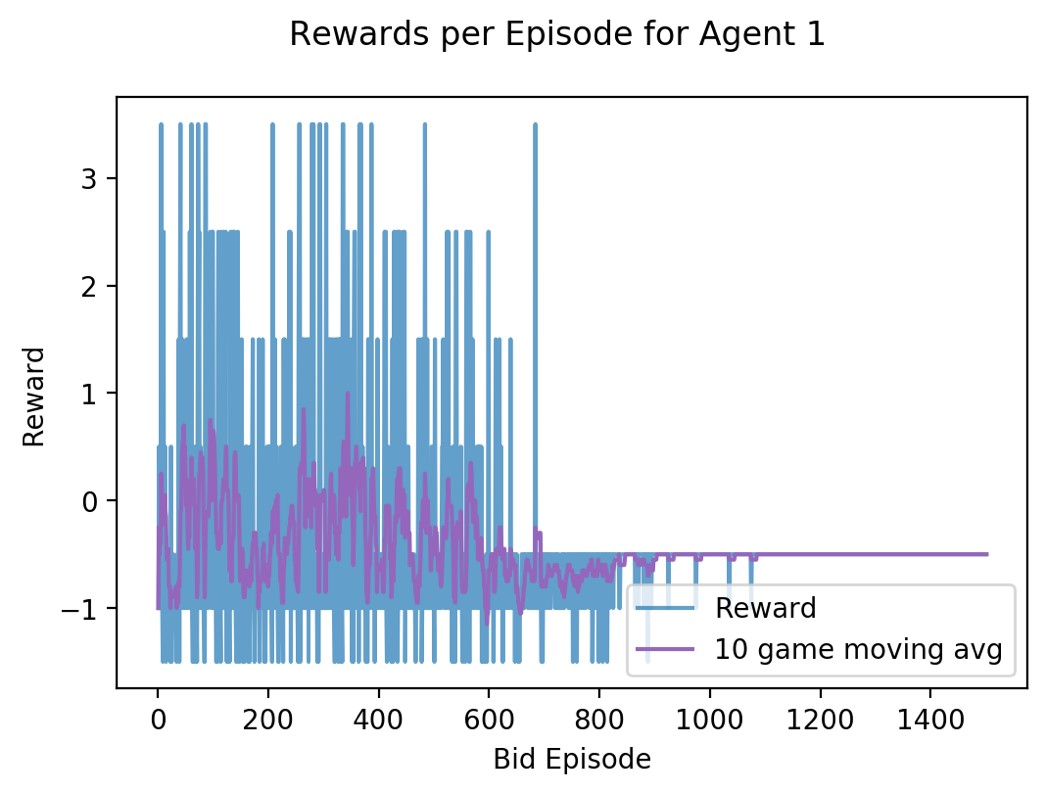}%
}

\subfloat[Reward function for the second player.]{%
  \includegraphics[clip,width=\columnwidth]{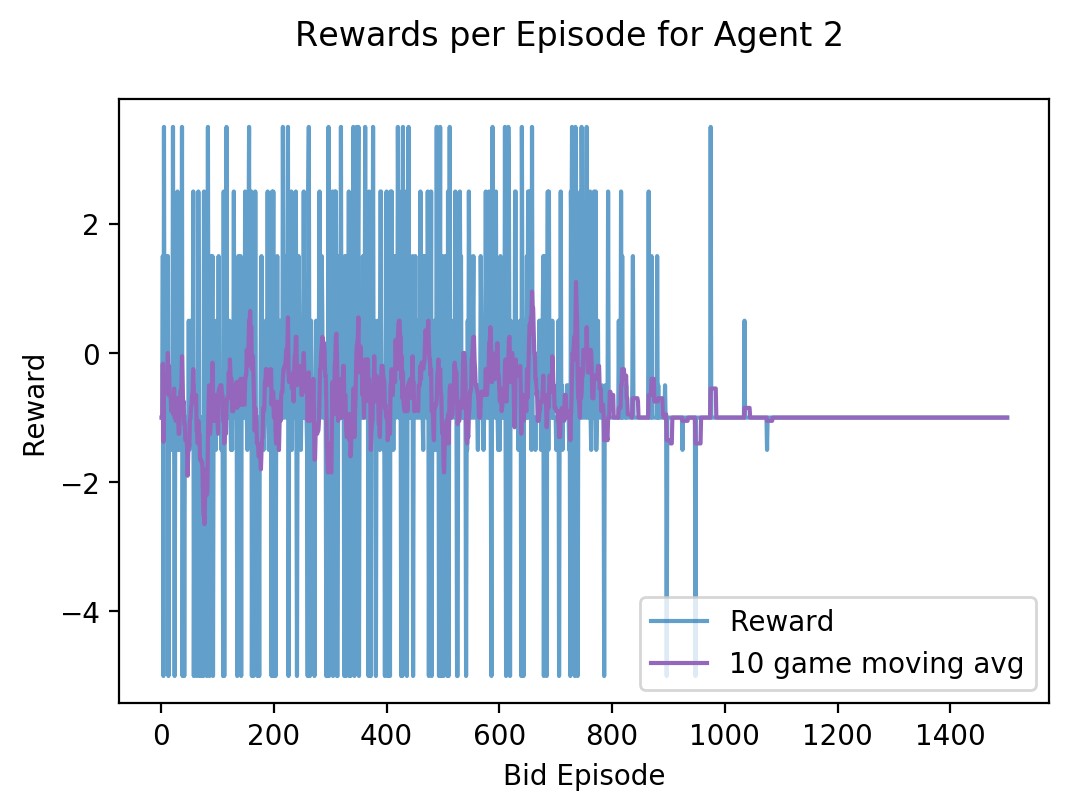}%
}

\caption{One Period Open Bid with a ceiling price of 5 for an indivisible good.}

\end{figure}

\begin{figure}[!t]
\centering
\includegraphics[width=3.2in]{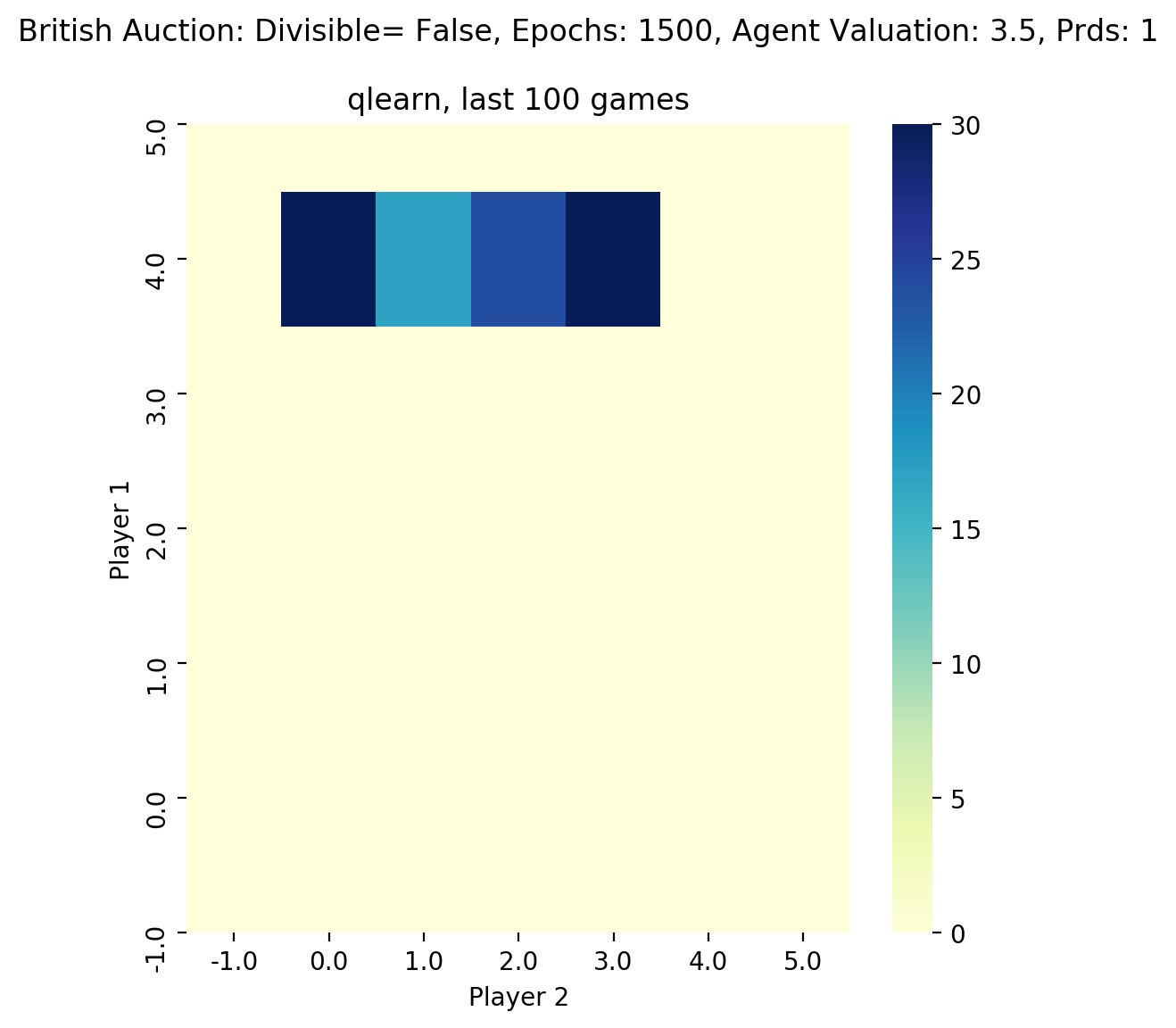}
% where an .eps filename suffix will be assumed under latex,
% and a .pdf suffix will be assumed for pdflatex; or what has been declared
\DeclareGraphicsExtensions.
\caption{Heat map representation of the last 100 games for two players.}
\label{fig3}
\end{figure}

As can be seen in Figure 2 and Figure 3, the second agent has some positive rewards in the initial stages (and around 1000 epochs as well) as a result of the “exploration” of the first agent to the environment. As the exploration parameters decay in the last 1000 epochs, the rewards for the agents become \$0.5 and -\$1, respectively. In  Figure 4 and Figure 5, the environment has been slightly extended to a higher ceiling price (\$20) while everything else remains constant. As can be seen from the graphs, the second agent has some positive rewards in the initial stages (and around 1000 epochs as well) as a result of the “exploration” of the first agent to the environment. As the exploration parameters decay in the last 1000 epochs, the rewards for the agents become \$0.5 and -\$1, respectively. In Figure 4 and Figure 5, the environment has been slightly extended to a higher ceiling price (\$20) while everything else remains constant. As can be seen from the graphs, the first agent always bids a slightly higher price (i.e., \$8) than his/her valuation, which covers the entry fee. On the other side, the second agent bids up to their valuation as well. The equilibrium points are said to be a Nash Equilibrium for this game as well.

\begin{figure}[htp]

\subfloat[Reward function for the first player.]{%
  \includegraphics[clip,width=\columnwidth]{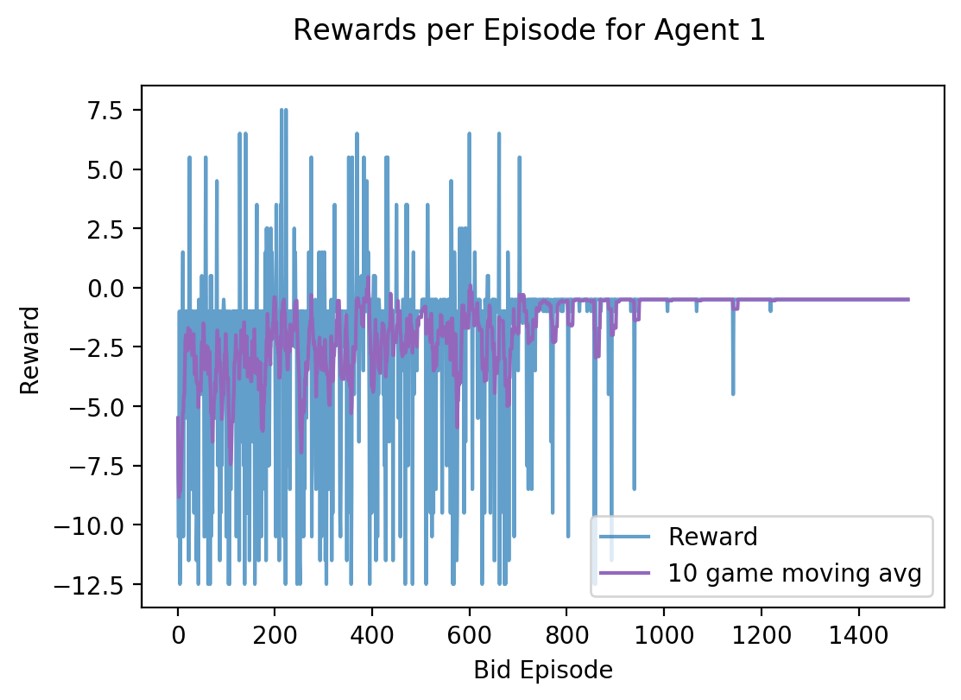}%
}

\subfloat[Reward function for the second player.]{%
  \includegraphics[clip,width=\columnwidth]{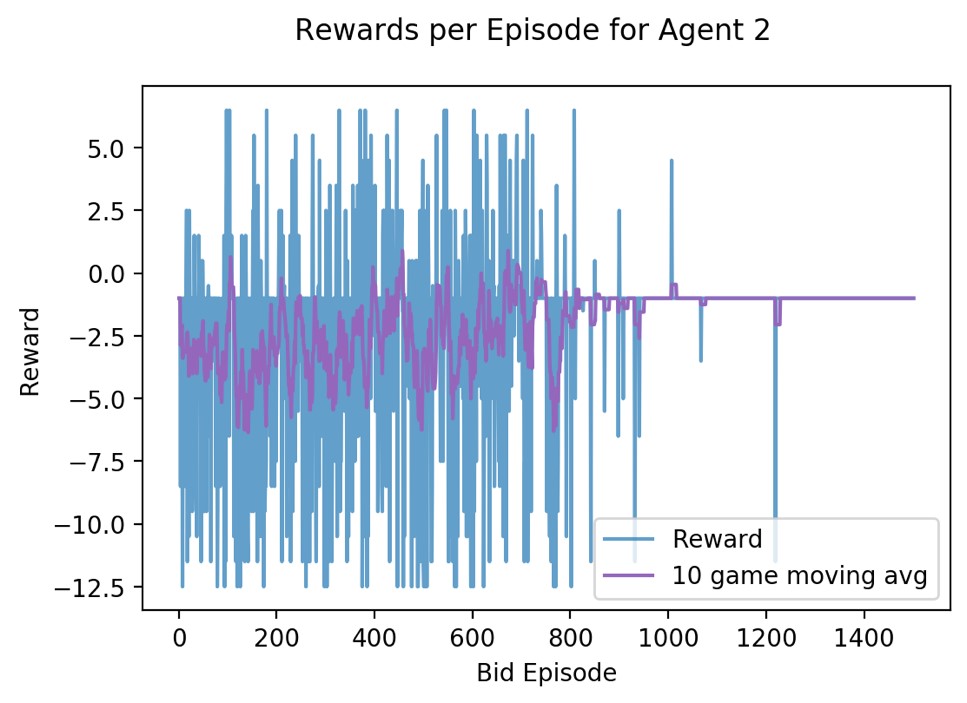}%
}

\caption{One Period Open Bid with a ceiling price of 20 for an indivisible good.}

\end{figure}

\begin{figure}[!t]
\centering
\includegraphics[width=3.2in]{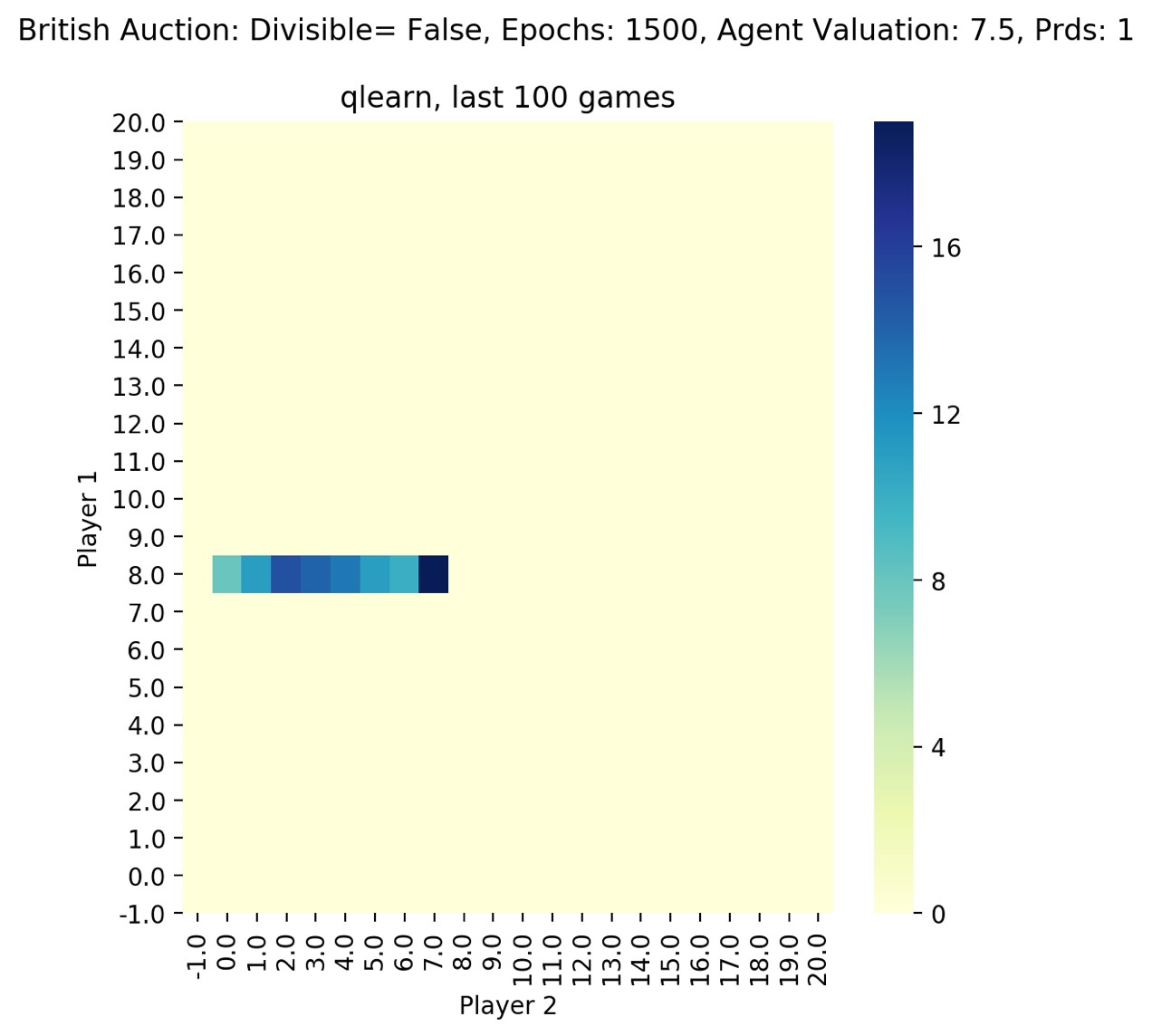}
% where an .eps filename suffix will be assumed under latex,
% and a .pdf suffix will be assumed for pdflatex; or what has been declared
\DeclareGraphicsExtensions.
\caption{Heat map representation of the last 100 games for two players.}
\label{fig5}
\end{figure}

This time, the type of auction is changed to the Close Auction. In this type of environment, RL agents do not see the opponent’s bids while deciding an action (bid) to take. The ceiling price is \$5, and there is also the entrance fee of \$1 paid by the losing player. The final decisions are built by the type of good, whether it is divisible or indivisible. Figure 6 shows the sealed (close) auction with an agent valuation of 3.5 of a divisible good.

\begin{figure}[htp]

\subfloat[Reward function for the first player.]{%
  \includegraphics[clip,width=\columnwidth]{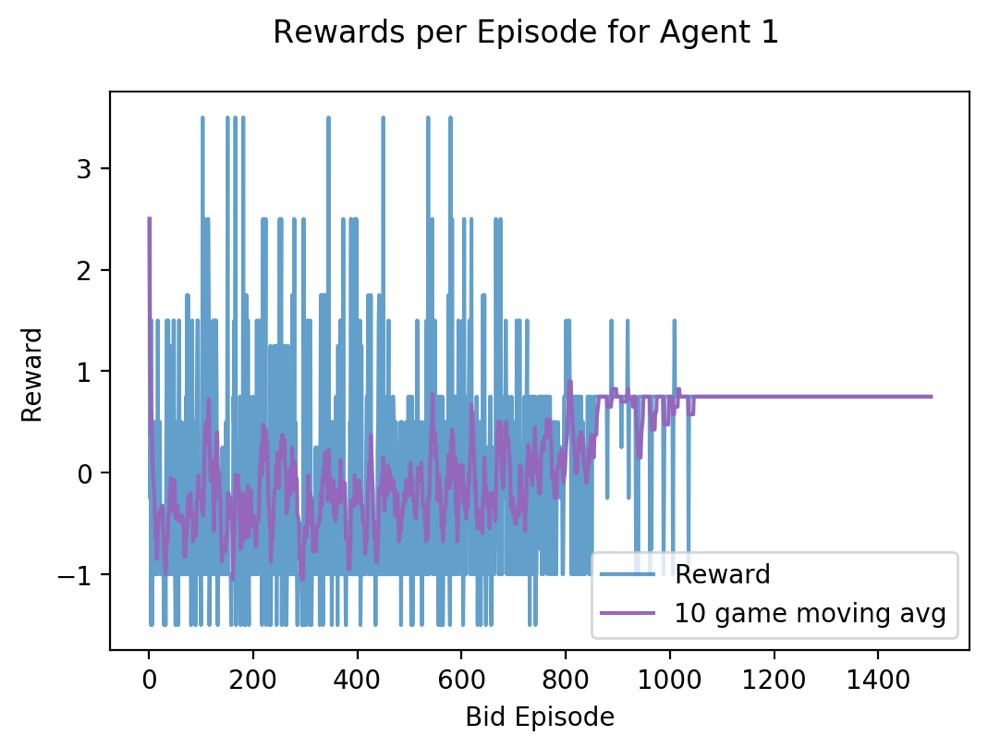}%
}

\subfloat[Reward function for the second player.]{%
  \includegraphics[clip,width=\columnwidth]{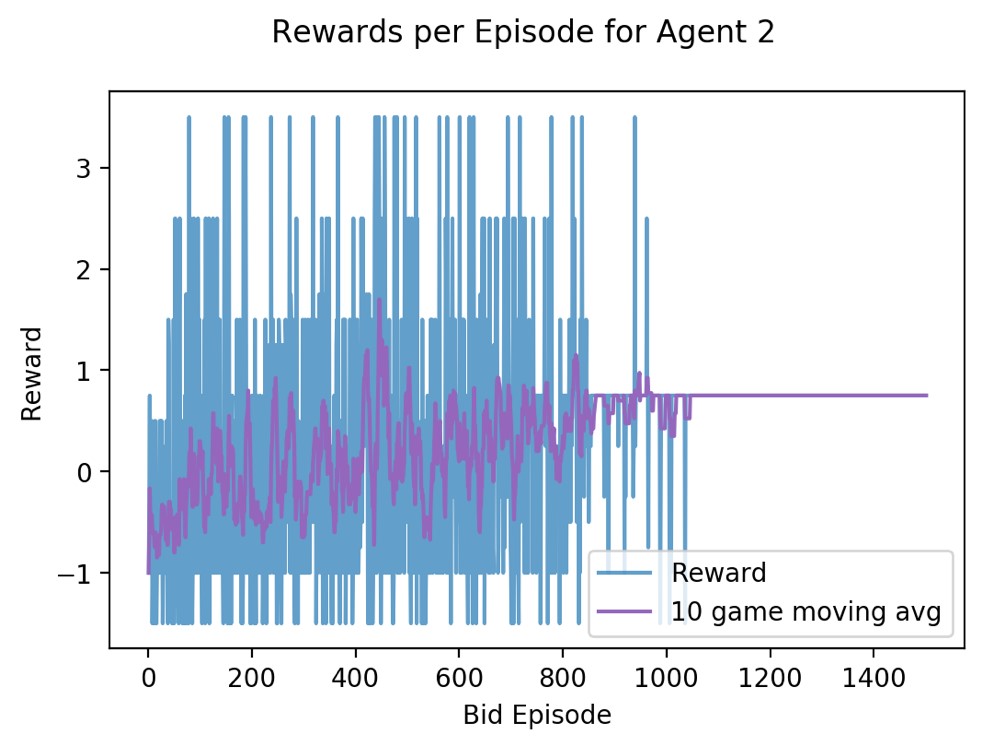}%
}

\subfloat[Heat map function.]{%
  \includegraphics[clip,width=\columnwidth]{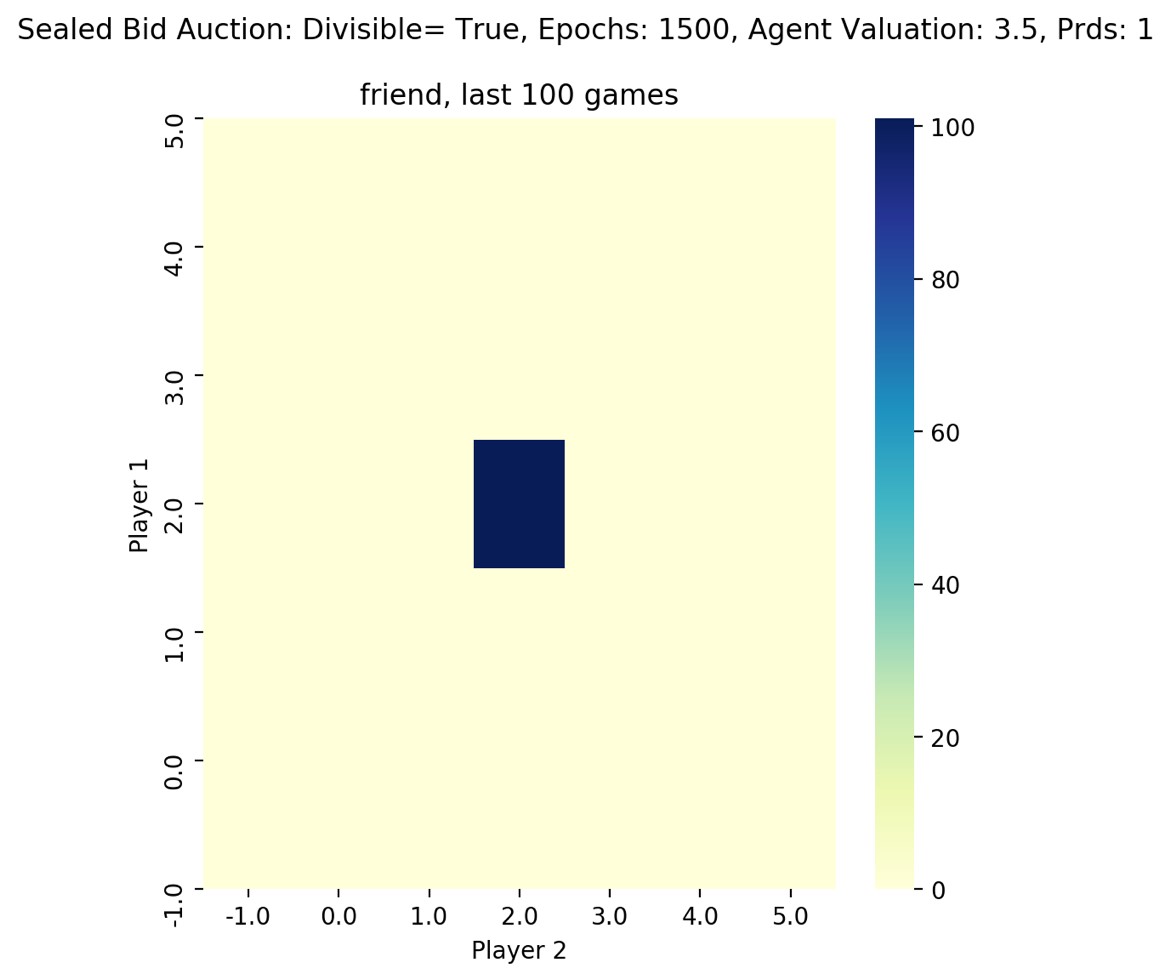}%
}

\caption{One Period Close Bid with a ceiling price of 5 for a divisible good.}

\end{figure}

As for the close bid for an indivisible good, both agents act as rational players in the game. The outcomes for both agents are negative (Figure 7). Since the good is indivisible, both the agents bid the amount of  \$4, and the winner is selected randomly. It is also worth to note that they are still in pure Nash Equilibrium point for the game.

\begin{figure}[htp]

\subfloat[Reward function for the first player.]{%
  \includegraphics[clip,width=\columnwidth]{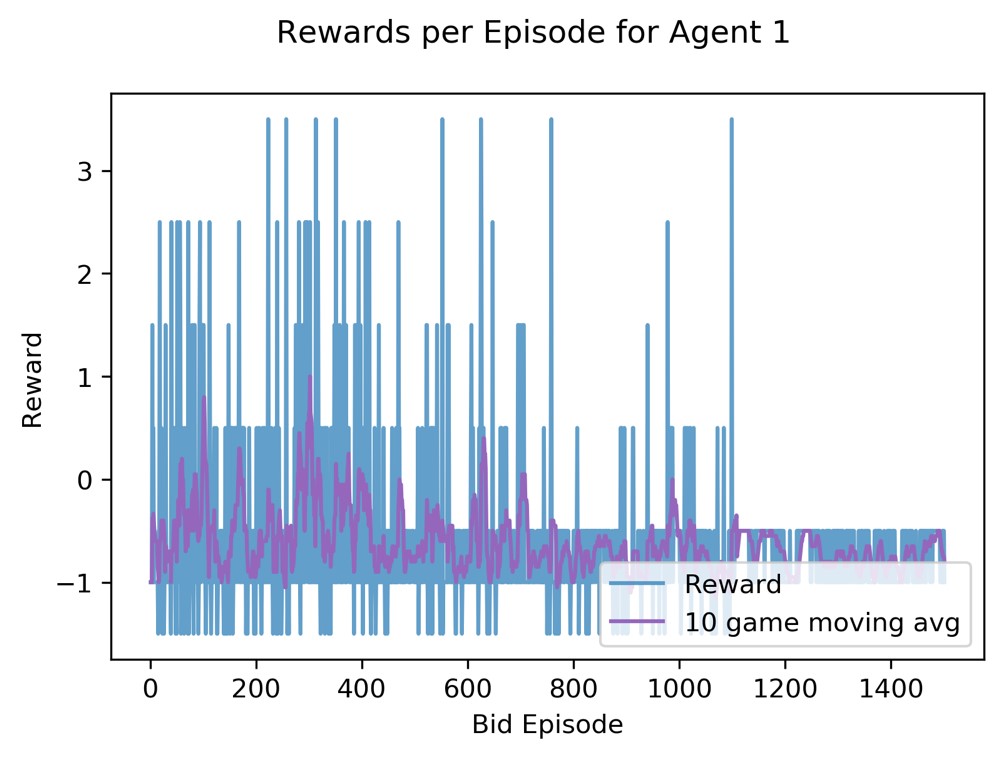}%
}

\subfloat[Reward function for the second player.]{%
  \includegraphics[clip,width=\columnwidth]{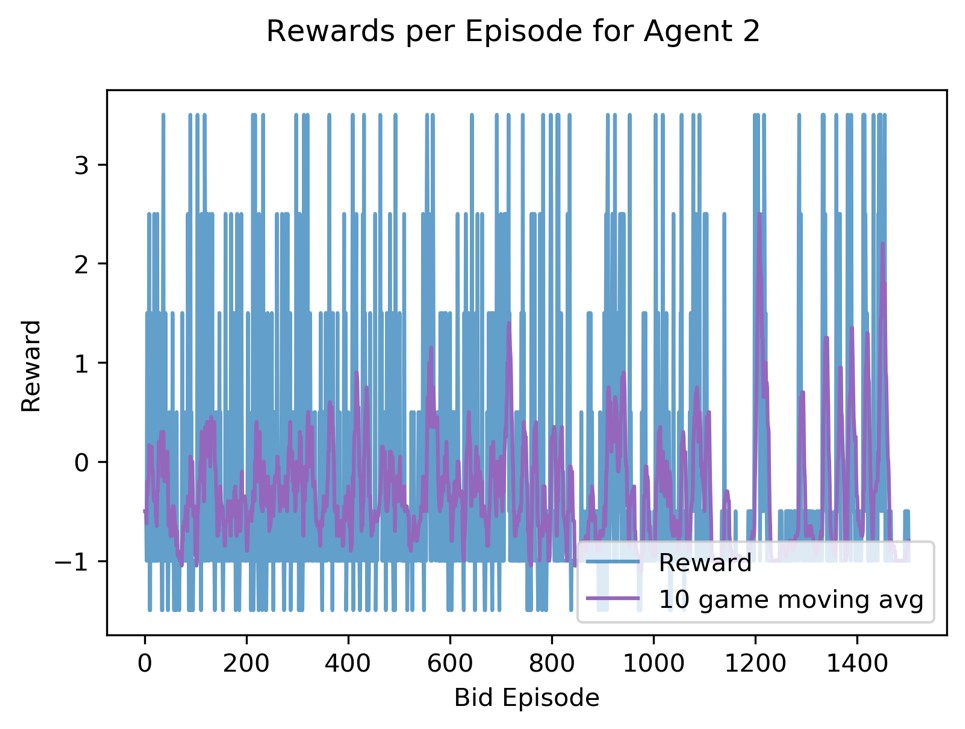}%
}

\subfloat[Heat map function.]{%
  \includegraphics[clip,width=\columnwidth]{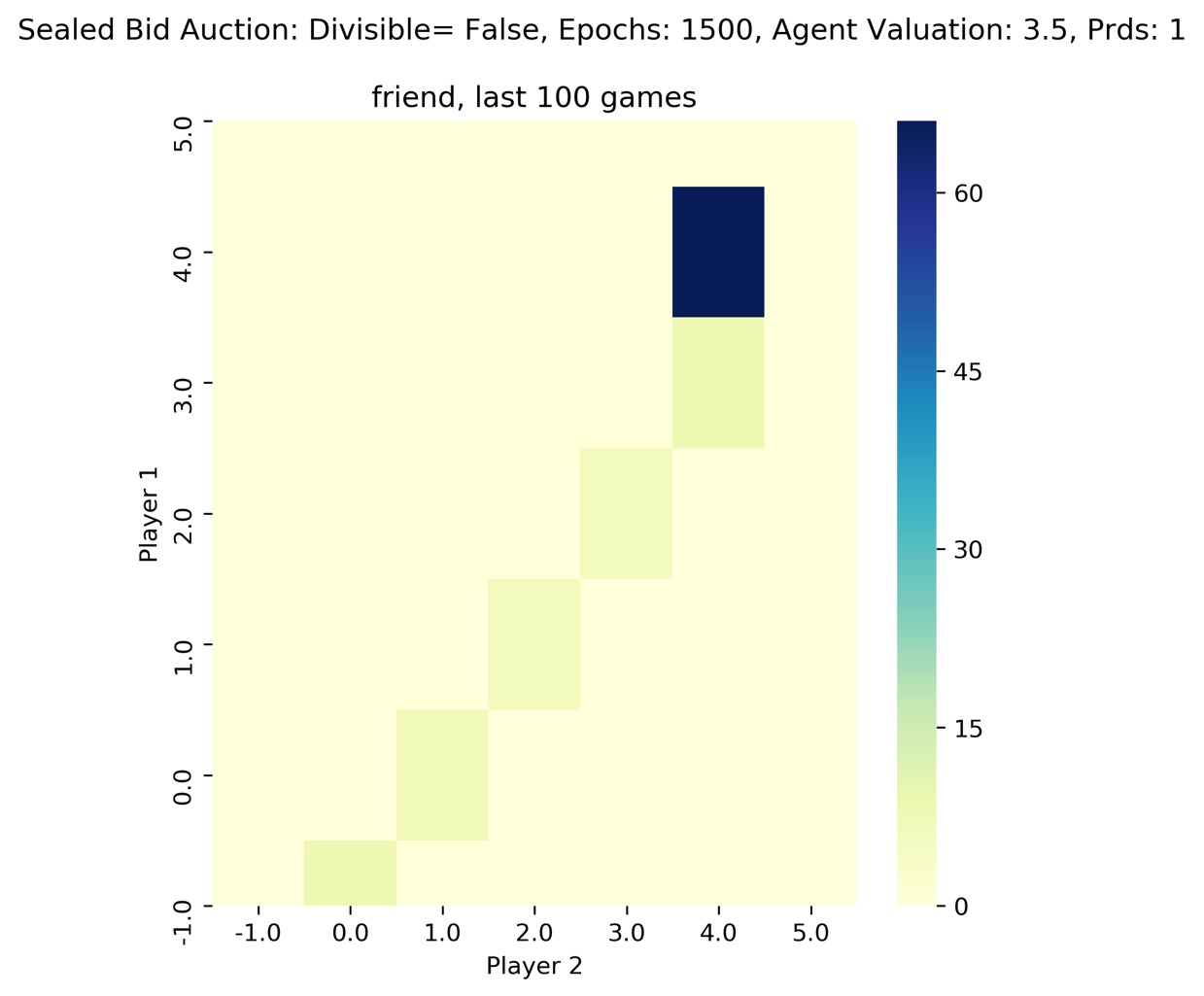}%
}

\caption{One Period Close Bid with a ceiling price of 5 for an indivisible good.}

\end{figure}

Let us suppose that the product is a divisible product and the agents can cooperate and collaborate for the product. The outcomes are summarized below (Figure 8).

\begin{figure}[!t]
\centering
\includegraphics[width=3.2in]{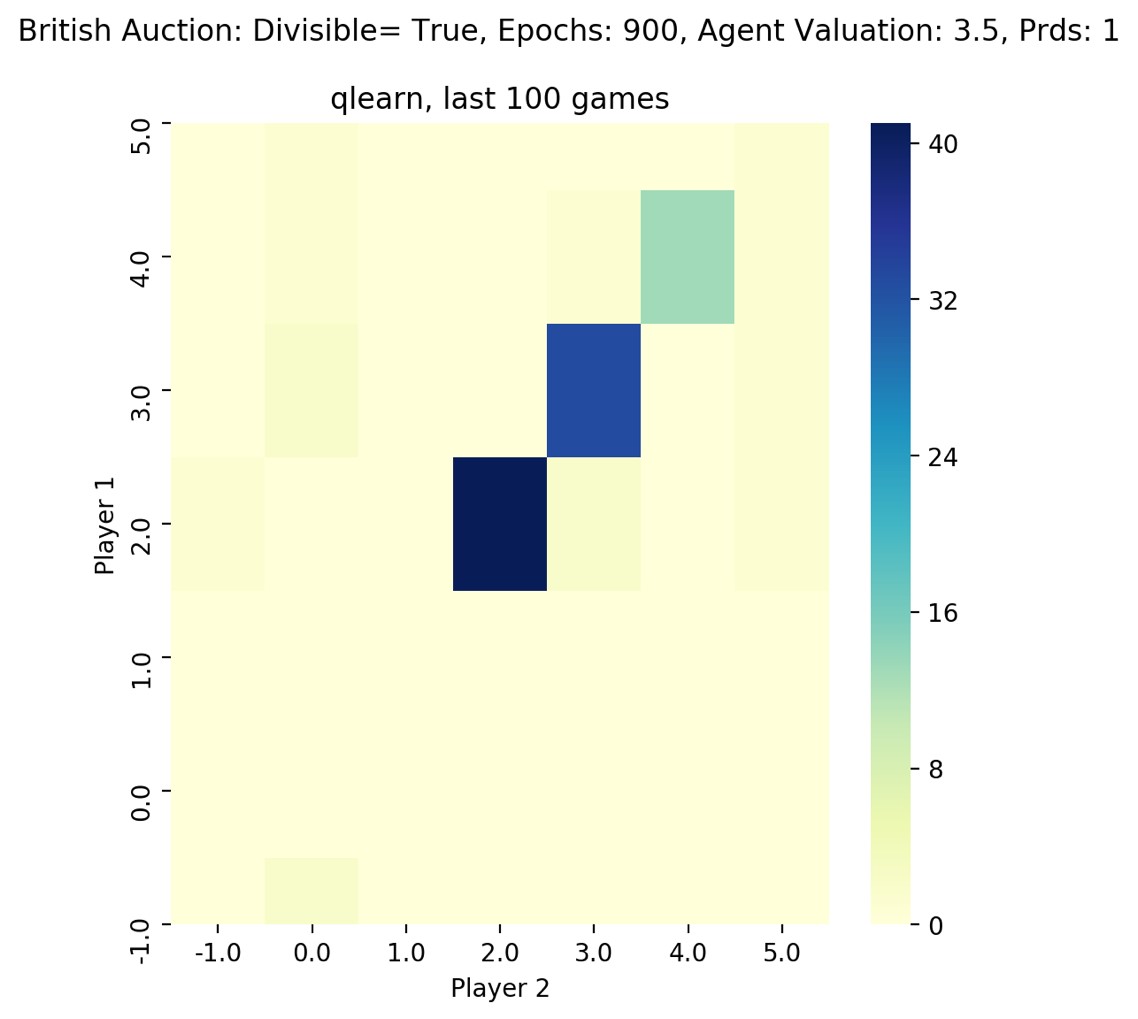}
% where an .eps filename suffix will be assumed under latex,
% and a .pdf suffix will be assumed for pdflatex; or what has been declared
\DeclareGraphicsExtensions.
\caption{Heat map functions for players, One period Open Bid for a divisible good.}
\label{fig8}
\end{figure}

The heat map for Figure 8 shows that 50\% of the games end in (2,2) and 20\% of the games end in (3,3) which yield positive outcomes for the RL agents and 30\% of the games end in (4,4) which gives negative outcomes to the RL agents. The points are still among the pure Nash solutions for the game. Figure 9 shows the Nash solutions for the same setup revealed in different types of agents (types: Friend and Foe).

\begin{figure}[htp]

\subfloat[Type of agents: Friend.]{%
  \includegraphics[clip,width=\columnwidth]{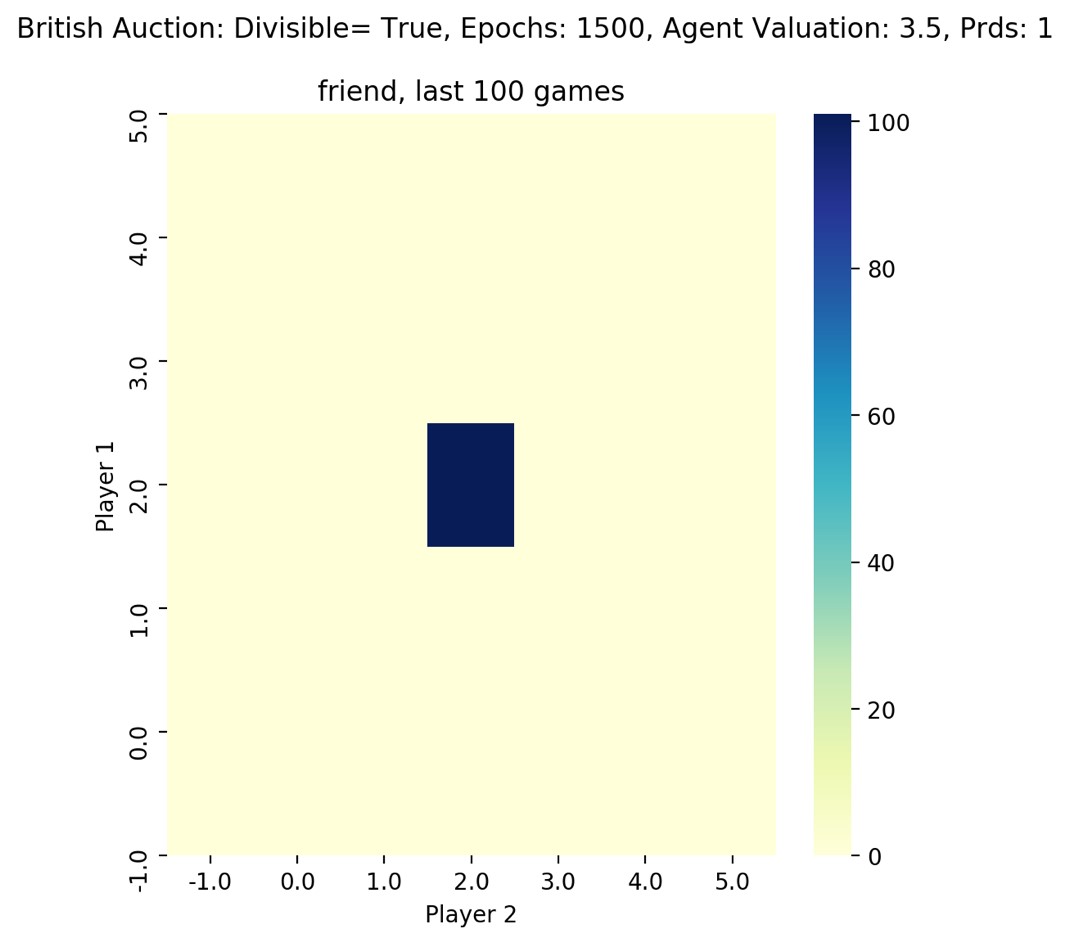}%
}

\subfloat[Type of agents: Foe.]{%
  \includegraphics[clip,width=\columnwidth]{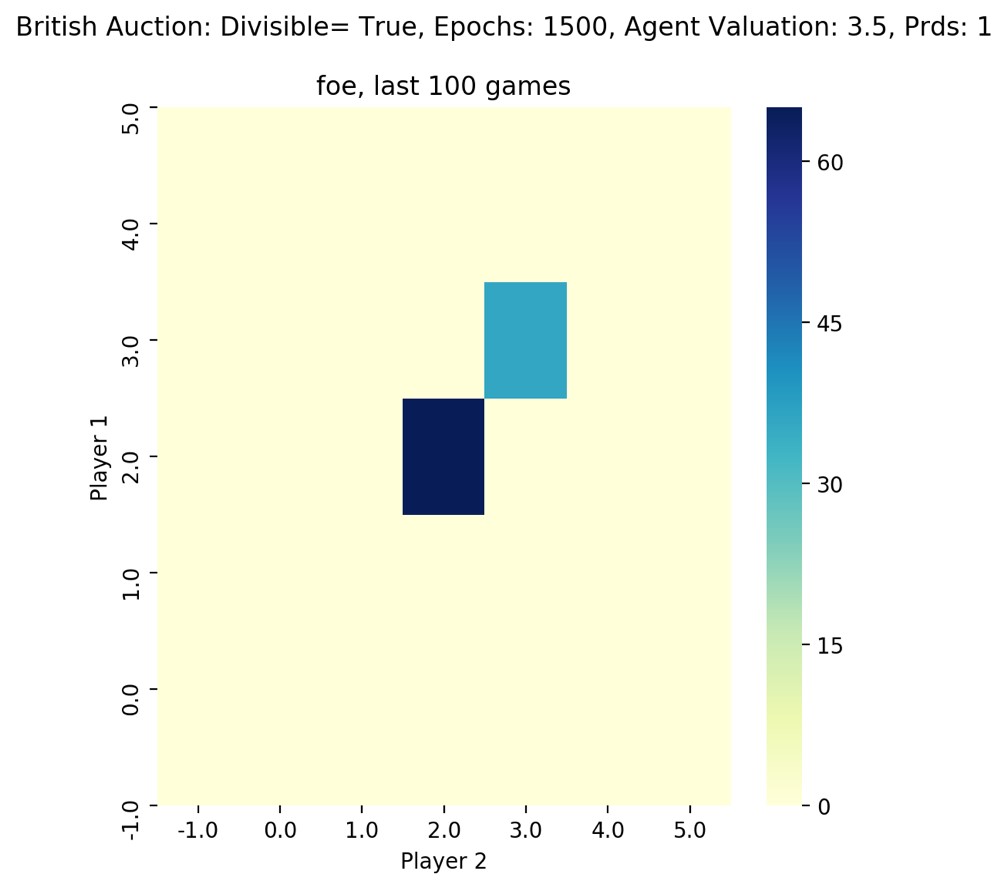}%
}

\caption{Heat map, Pure Nash Equilibria in Different Types of Learners.}

\end{figure}

As for multi-period bidding, the RL agents are supposed to play with a chance to bid more than one bidding. This is the environment in which each of the agents has more turns as long as the auction continues. For this reason, the setup for the game was changed so that each player can have one more chance to bid (two periods). Here is the new setup for the RL agents: Two periods, divisible good, and the base price is \$3.5 for both of the agents. As can be seen in Figure 10, the first agent sticks to the bid \$4 at the very beginning, and the second agent draws the game. Although this bid returns a negative outcome to the first bidder, choosing a different bid yields more negative outcomes.

\begin{figure}[htp]

\subfloat[Bids per period for the first player.]{%
  \includegraphics[clip,width=\columnwidth]{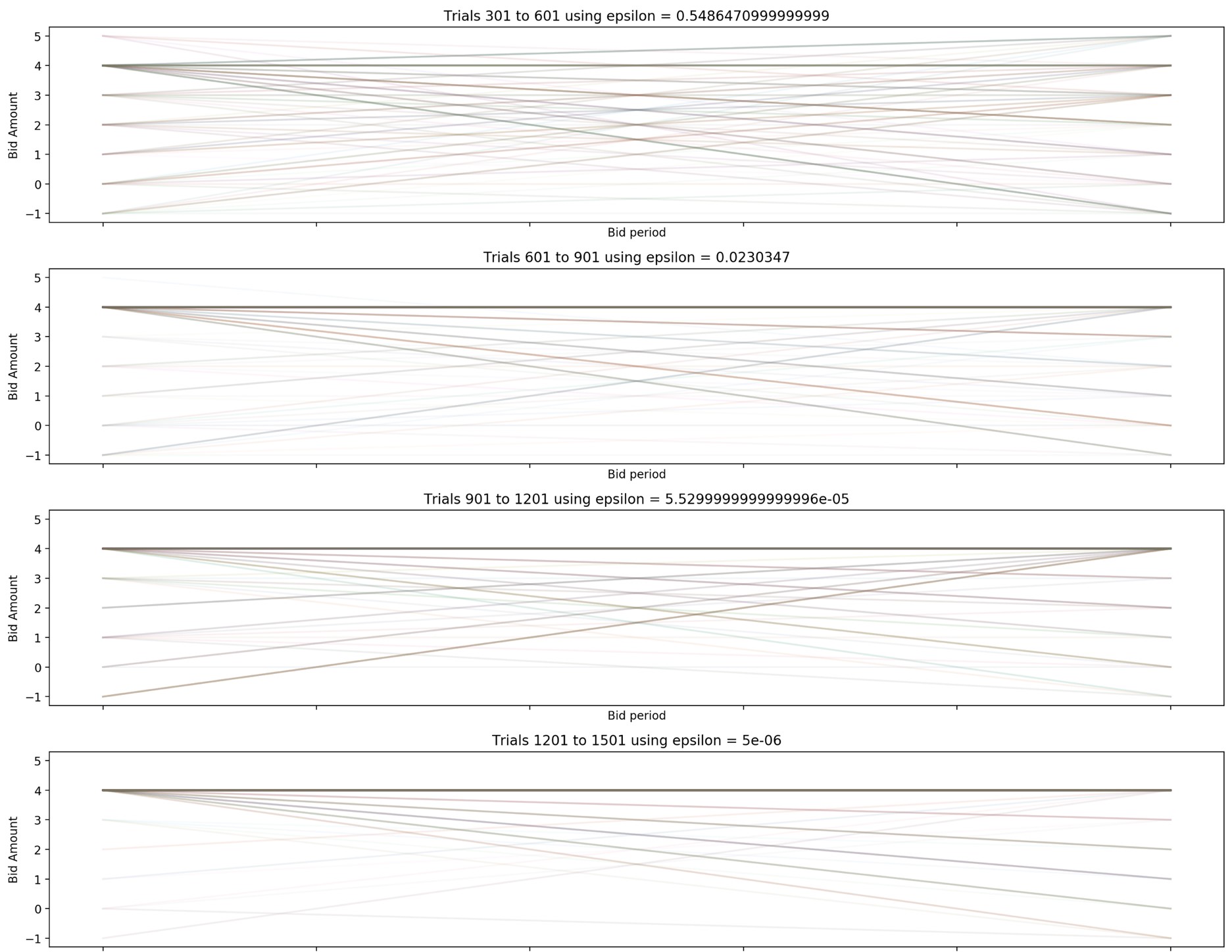}%
}

\subfloat[Reward function of the first player.]{%
  \includegraphics[clip,width=\columnwidth]{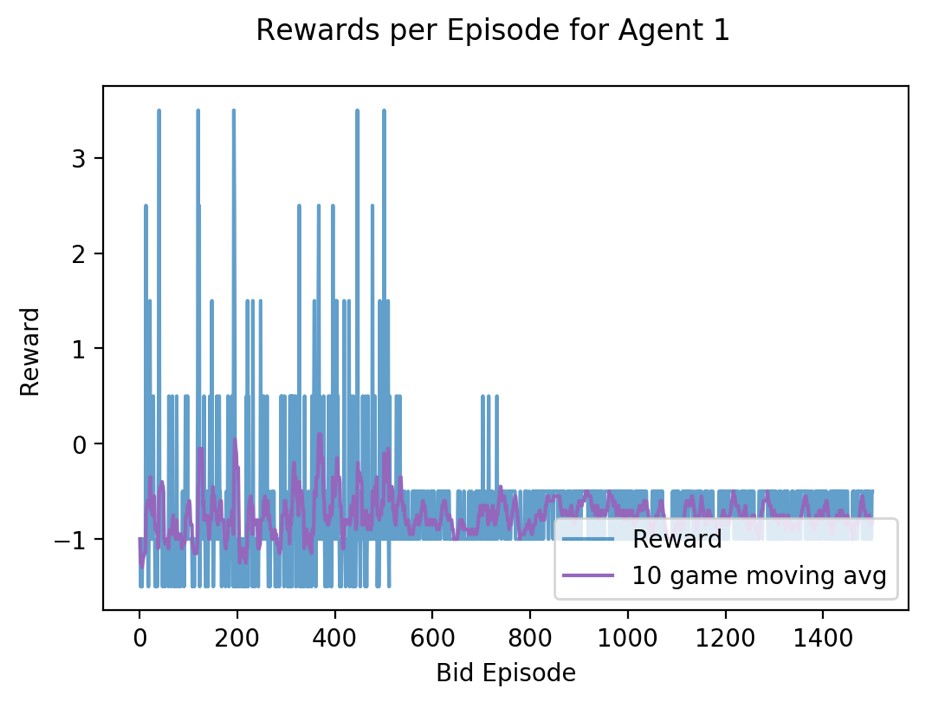}%
}

\subfloat[Reward function of the second player.]{%
  \includegraphics[clip,width=\columnwidth]{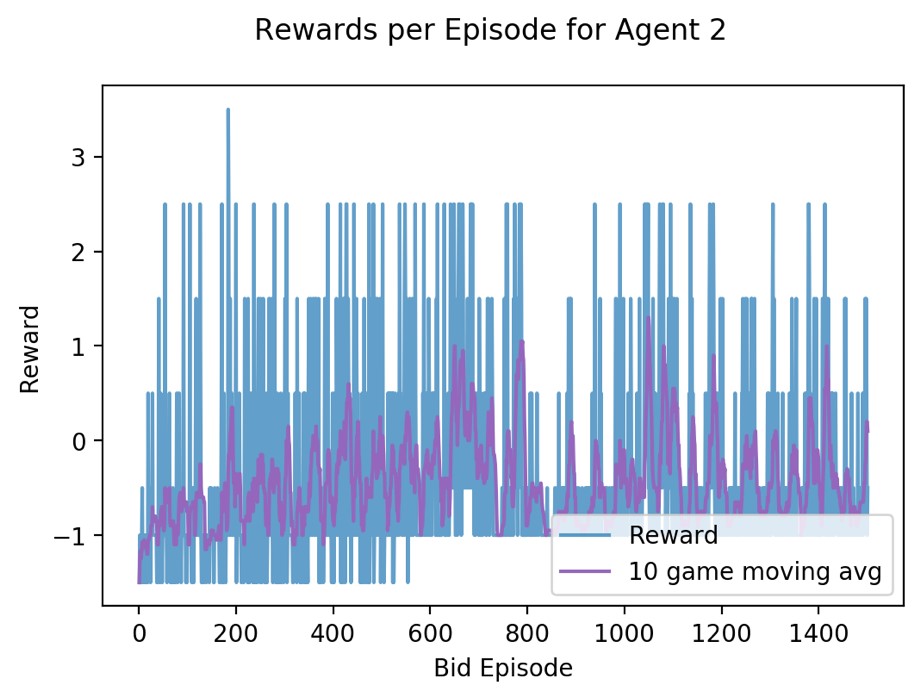}%
}

\caption{Reward functions for players and the multi-period bidding for a non-divisible good.}

\end{figure}

The last part of the auction design is about Vickrey Auction. In Vickrey Auction, the winner is the bidder who pays the highest, and the winner price may differ: it may be the second-best, or third, even the minimum bid. In this section, it is supposed that the winner is the agent who has the highest bid, and the winner price is the minimum bid, which is the loser agent’s bid. In this game setup, the valuation is \$3.5 for both agents, the loser pays \$1, and if there is a case of the same bid, both agents pay \$1.5 for re-setup. As can be seen in Figure 11, the first agent decides to bid always the highest bid (\$5), which means that the agent has a dominant strategy, but the second agent does not bid the highest bid which results in a negative outcome (-\$1.5). Thus, any bid lower than \$5 results in a case that the first agent gets the product with the second agent’s bidding price. It is worth to note that there is no dominant strategy for the second agent.

\begin{figure}[htp]

\subfloat[Heat map for the game, valuation 3.5.]{%
  \includegraphics[clip,width=\columnwidth]{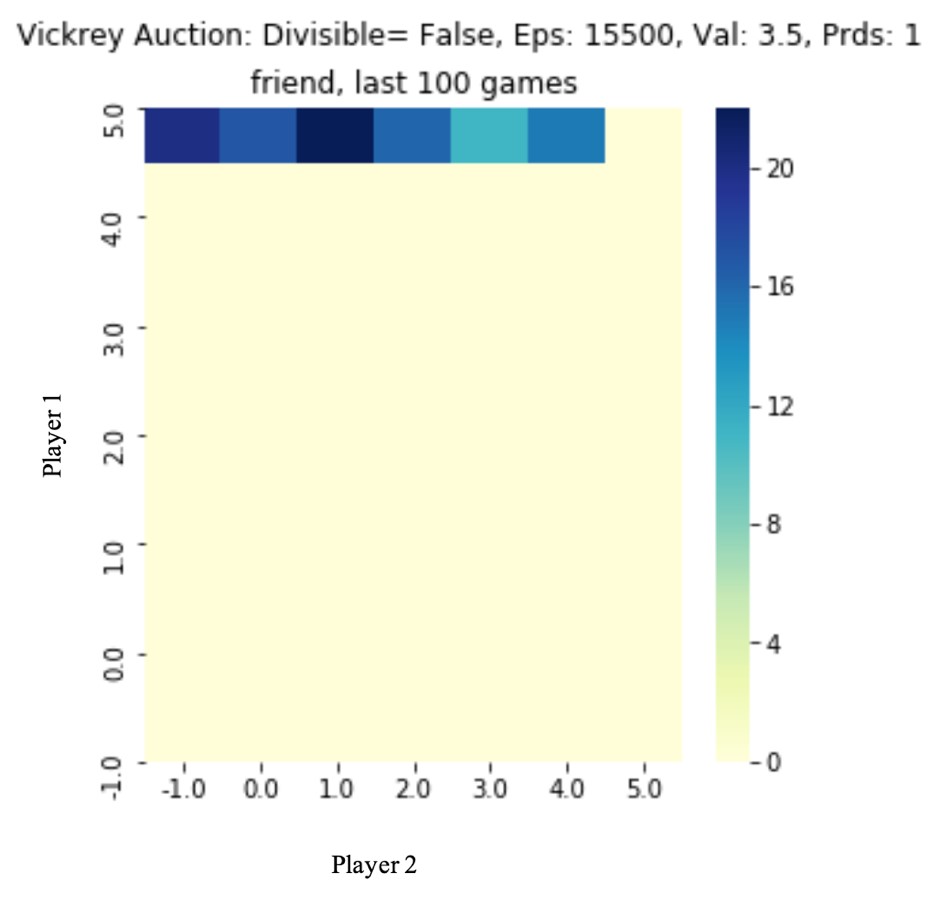}%
}

\subfloat[Reward function of the first player.]{%
  \includegraphics[clip,width=\columnwidth]{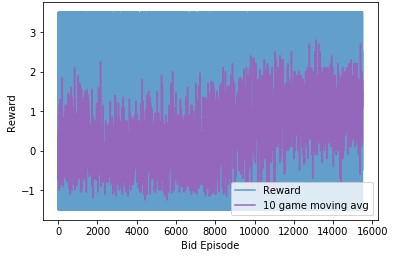}%
}

\subfloat[Reward function of the second player.]{%
  \includegraphics[clip,width=\columnwidth]{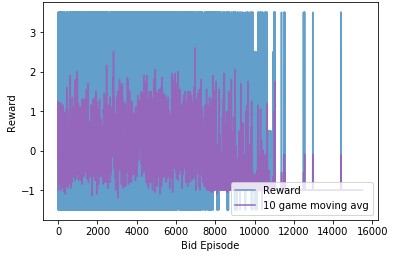}%
}

\caption{Reward functions of the players during Vickrey Auction with non-divisible goods.}

\end{figure}

This time, the good is changed to divisible ones. If both agents agree on a price, they will get half of the product, and pay half of the price they bid. This time, both agents bid \$4, and get a negative outcome (-\$0.25). It is worth to note that there is a chance for both agents to bid lower than their true valuation and get a positive outcome, but on the contrary, their bids are slightly higher than their true prices (\$3.5). The MARL agents show that the Vickrey auction reveals the true valuation of the agents.

If the true values of the agents are changed to 3, it is expected that both agents will give their true valuation as their bids (entrance fee is \$1). Figure 12 and Figure 13 below illustrate the Vickrey Auction for both divisible and non-divisible goods. It can be seen that the RL agents bid their true valuation for the product. The agents act as individually rational and in a truthful manner. The result which they decide is still a Nash equilibrium for the game.

\begin{figure}[htp]

\subfloat[Reward function of the first player.]{%
  \includegraphics[clip,width=\columnwidth]{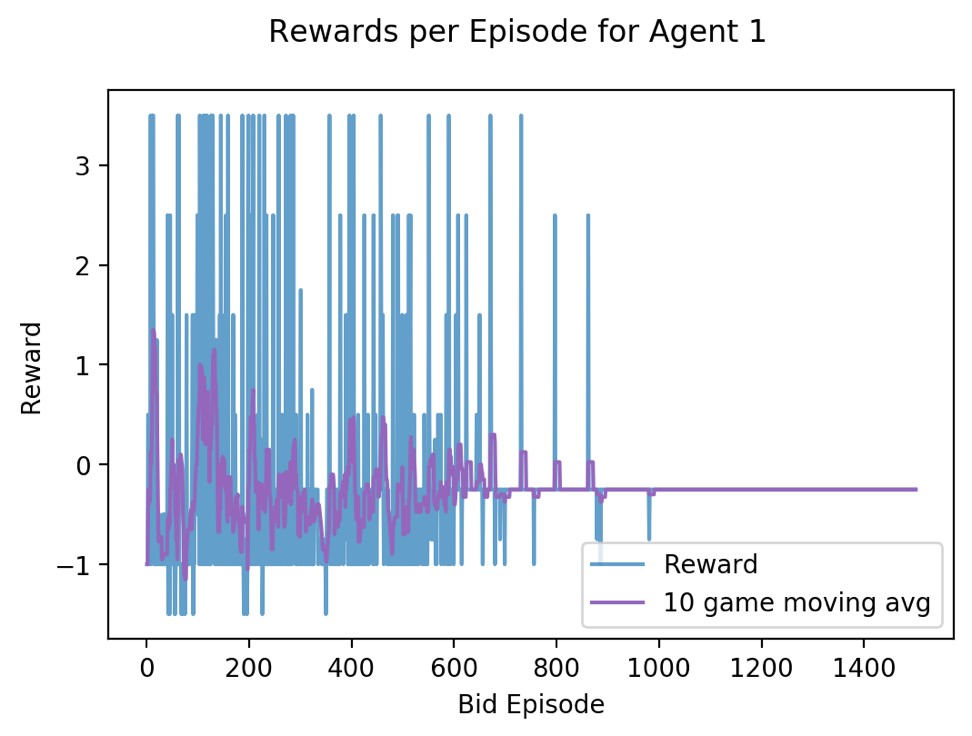}%
}

\subfloat[Reward function of the second player.]{%
  \includegraphics[clip,width=\columnwidth]{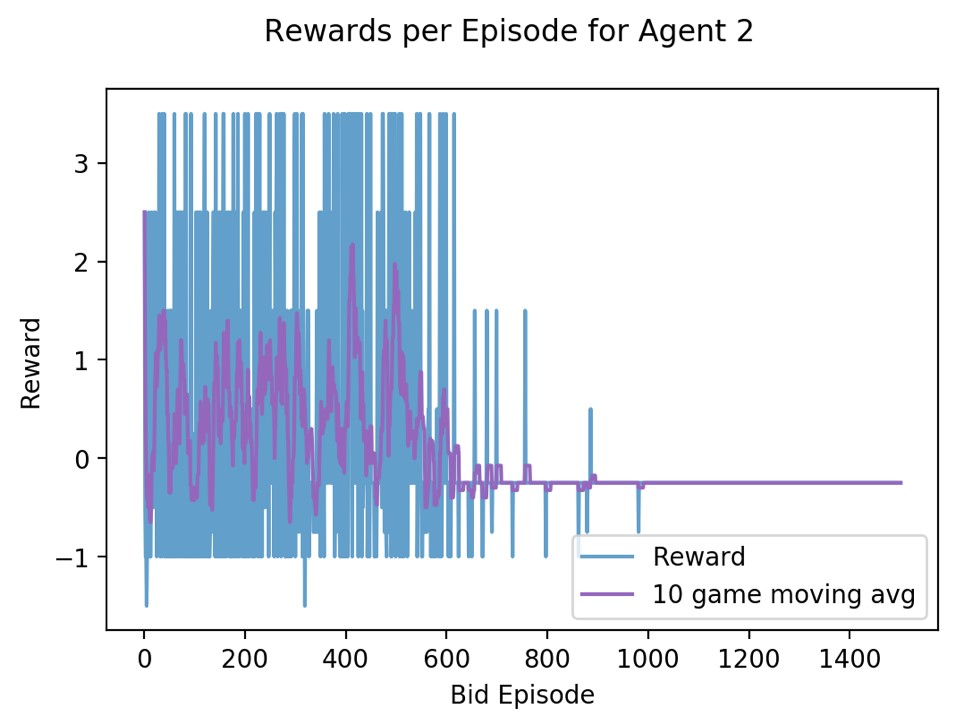}%
}

\caption{Reward functions for players, Vickrey Auction with non-divisible goods.}

\end{figure}

\begin{figure}[htp]

\subfloat[Heat map for valuation of 3.5.]{%
  \includegraphics[clip,width=\columnwidth]{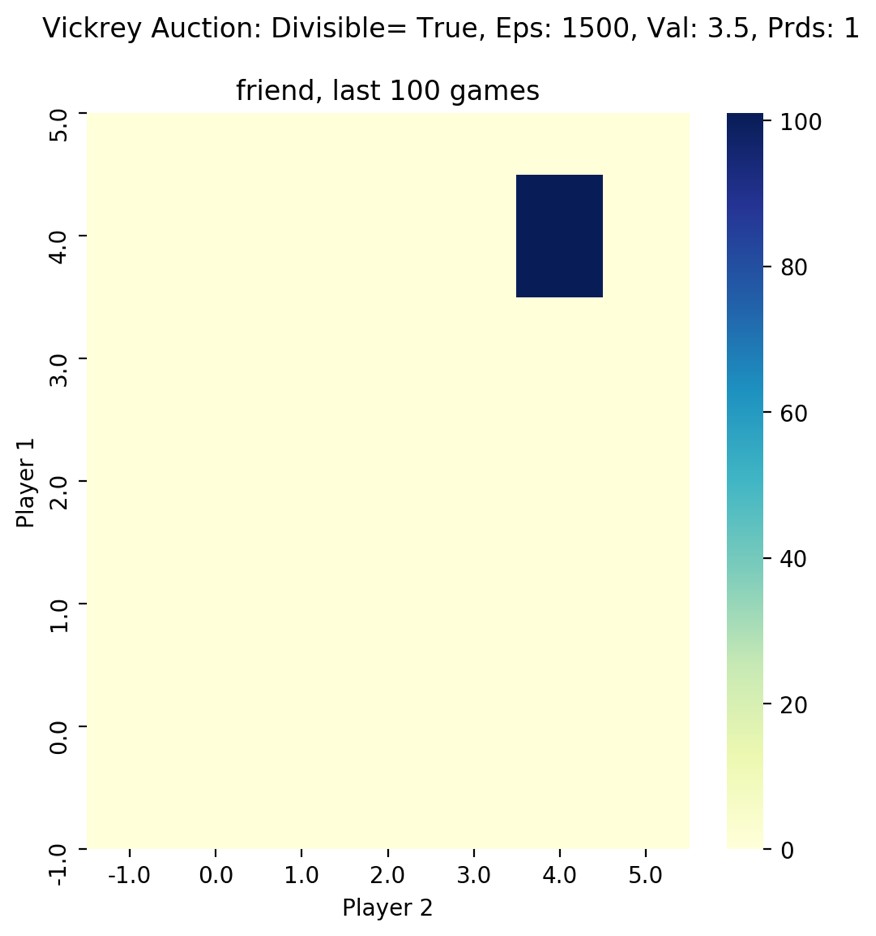}%
}

\subfloat[Heat map for valuation of 3.0.]{%
  \includegraphics[clip,width=\columnwidth]{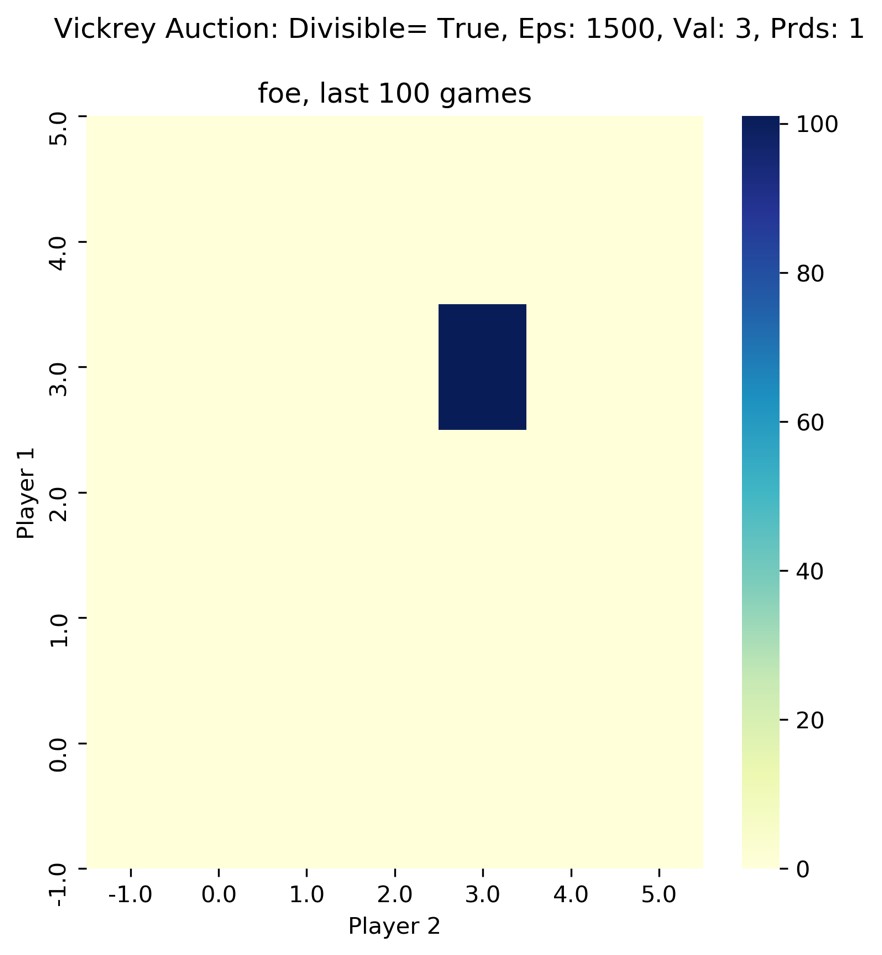}%
}

\caption{Heat map values, Vickrey Auction with non-divisible goods.}

\end{figure}

\section{Discussion and Conclusion} \label{sec:discussion}

In this paper, it is questioned whether the agents created by RL algorithms are the rational agents in different types of auction mechanisms or not. The equilibrium points in different types of bidding environments determined by these agents show that they act as rational agents. The bidding strategy of these agents in each type of auction mechanism revealed the fact that the RL agents are individually rational and strategy-proof. The strategic agents created by the RL algorithms are supposed to be the bidder agents in different types of auction mechanisms such as British, Sealed Bid Auction, and Vickrey Auctions mechanisms.

The first experiment is about the decision mechanism of RL agents in British auction design. In this scenario, one period-British auction model for a divisible good, each agent having just one chance to bid and the order of bidding is sequential: the second agent sees the first agent’s price (open bid) and decides on bidding or not. The results show that RL agents play individually rational in this type of environment, so the outcome of the British auction yields a Nash equilibrium.

The second experiment is about sealed bid auction (close bid). In this type of auction, RL agents cannot see the other’s bid while deciding on an action. The joint action of the RL agents determines the winner of the game. In this game setup, both agents act as rational players as well. Thus, the outcomes for both agents yield negative results. It is also worth noting that RL agents are still in pure Nash Equilibrium of the game.

As for an indivisible good, both agents act as rational, too. Both RL agents bid the same price (i.e., \$4), so the outcomes for both agents become negative. Since the good is indivisible and both the agents bid the same amount, the winner is selected by picking one of them randomly. It is also worth noting that the state where the game ends is still the pure Nash Equilibrium point of the game.

As far as a divisible good in an open bid, we assume that the agents can collaborate for the product. In this type of game design, the states where the game ends that are constituted as different points in the game. That is to say, 50\% of the games end in (\$2, \$2) and 20\% of the games ends in (\$3, \$3) which yield positive outcomes for the RL agents and 30\% of the games end in (\$4, \$4) which gives negative outcomes to the agents. The points are still among the pure Nash solutions for the game. Similarly, when the learning algorithm is changed from Q-Learning to Foe and Friend, the ending states are also changed. In Foe, the ending state is (\$4, \$4), which brings negative outcomes for the agents, whereas in Friend, the ending state is (\$1, \$1), which yields the highest possible positive outcome for the game. Thus, we find that as the learning algorithm changes, the ending states for the games also do change.

From a single-stage game, we turned to multiple stages for the games where the RL agents are supposed to play with a chance to bid more than one bidding. In this type of auction environment, both agents have more turns unless the game ends. The results show that the game ends at the beginning stage, where it is a Nash equilibrium of the game, even though the outcome is negative for the agents.

Lastly, we conduct a Vickrey Auction design for the indivisible product. Although the game has many pure Nash Equilibrium states, the game ends with the states where one of the RL agents has a dominant strategy of \$5, and the other agent has no dominant strategy, changing from “not enter” to \$4. Since the product is indivisible and the loser gets a negative outcome, and any bid lower than \$5 results in a weakly less negative outcome for the first agent. Thus, the agent who has a dominant strategy (\$5), gets the product with the second agent’s bidding price. It is worth saying that if one of the agents insists on the same strategy over the others, the second agent adjusts the situation and differs its strategy. In other words, in the case of paying \$1.5 for readjusting the auction environment, the agent decides to lose \$1 with bidding different than \$5. In this Vickrey auction, the game ends with the highest possible winning bid, \$5, but the winner has to pay the bid of the second agent, which is less than \$5. Therefore, the agent who has a dominant strategy gets both positive and negative outcomes depending on the second agent’s biddings. Even though reaching the pure Nash equilibrium points is not seen possible by iteration of the strategies of the agents, the RL agents learn to act as a rational and strategy-proof player and reach the Nash equilibrium. As for Vickrey auction for divisible goods, both agents have the same strategy to bid (\$4) and get a negative outcome (-\$0.25). It is worth noting that both agents choose to bid consistent with the appraisal for the product (i.e., \$3.5), which is the main finding of this game setup. Even though there is room for both agents to bid lower than their true valuation and get a positive outcome, both of the RL agents bid with their true prices, which means the Vickrey auction reveals the true valuation of the agents.

In this paper, multi-agent reinforcement learning agents in different auction setups were created. At the initial state of the game, both agents have no experience with the type and the mechanism of the auction. With a reward function and the state information, both RL agents become rational and strategy-proof bidding agents. It is also worth noting that the final state of the RL agent in the Vickrey auction for divisible goods reveals the fact that Vickrey Auction also gives similar solutions to the bidders compared to the other type of the auction mechanism. For this reason, the RL agents we create in this study can also be used to analyze how a rational and strategy-proof agent acts in a newly created auction environment. Also, in the Vickrey auction for indivisible goods, we did not expect to see a dominant strategy for an agent since both agents have the same hyperparameters for the game. The reason why they are not identical agents is that they differ in the “exploration” phase of the game setup. In other words, since both agents explore the environment on a random scale, they learn different information within the same state since their actions may differ. Even though we use the same learning rate and epsilon value for the RL agents, they become different agents. For this reason, we do not use different parameters for the agents, which can be done in further research. On the other hand, in real life, some agents are risk-averse, and some of them are greedy. Therefore, mimicking this type of human interaction in the auction environment by RL agents would be interesting and resourceful.

\end{document}